\documentclass[usegraphicx,useAMS,usenatbib]{mn2e}

\newcommand{\uflux}{{\rm erg}\, {\rm cm}^{-2}\, {\rm s}^{-1}}

\title[Warm absorber of H1419+480]{The warm absorber of the Seyfert 1 galaxy H1419+480}
\author[Barcons, Carrera \& Ceballos]
       {X. Barcons\thanks{E-mail: barcons@ifca.unican.es},
	 F.J. Carrera, M.T. Ceballos  \\
Instituto de F\'\i sica de Cantabria (CSIC-UC), 39005 Santander, Spain}

\date{June  2003}

\pagerange{\pageref{firstpage}--\pageref{lastpage}}
\pubyear{2003}

\begin{document}

\maketitle

\label{firstpage}

\begin{abstract}
The bright Seyfert 1 galaxy H1419+480 ($z\sim 0.072$), whose X-ray
colours from earlier HEAO-1 and $ROSAT$ missions suggested a complex
X-ray spectrum, has been observed with {\it XMM-Newton}.  The EPIC
spectrum above 2 keV is well fit by a power-law with photon index
$\Gamma=1.84\pm 0.01$ and an Fe K$\alpha$ line of equivalent width
$\sim 250$~eV. At softer energies, a decrement with respect to this
model extending from 0.5 to 1 keV is clearly detected. After trying a
number of models, we find that the best fit corresponds to OVII
absorption at the emission redshift, plus a 2$\sigma$ detection of
OVIII absorption.  A photoionised gas model fit yields $\log\xi\sim
1.15-1.30$ ($\xi$ in ${\rm erg}\, {\rm cm}\, {\rm s}^{-1}$) with
$N_H\sim 5\times 10^{21}\, {\rm cm}^{-2}$ for solar abundances.  We
find that the ionized absorber was weaker or absent in an earlier
$ROSAT$ observation. An {\it IUE} spectrum of this source obtained two
decades before shows a variable (within a year) CIV
absorber outflowing with a velocity $\sim 1800\, {\rm km}\, {\rm
s}^{-1}$. We show that both X-ray and UV absorptions are consistent
with arising in the same gas, with varying ionisation.

\end{abstract}


\section{Introduction}

X-ray spectral studies of radio-quiet Active Galactic Nuclei (AGNs),
reveal very frequently the presence of ionised material. This was first
noted by Halpern (1984), based on {\it Einstein} IPC data, where it
was also noted that these absorbers vary.  Evidence for ionized
absorption along the line of sight was provided by Nandra \& Pounds
(1992), based on the detection of an absorption edge in $ROSAT$
observations of the Seyfert 1 galaxy MCG-6-30-15. Fabian et al (1994)
were able to disentangle the OVII and OVIII K-edges of this warm
absorber using the much better spectral resolution of $ASCA$. 

In the years that followed, it became evident that warm absorbers are
ubiquitous among Seyfert 1 galaxies.  Reynolds (1997) and George et al
(1998) showed that more than 50\% of Seyfert 1s contain a warm
absorber.  These are produced by partially ionised gas with total
column density in the range $N_H\sim 10^{21}-10^{23}\, {\rm cm}^{-2}$,
probably located at or outside the Broad Line Region 
(Reynolds \& Fabian 1995).

With the advent of higher resolution grating spectrographs on
board {\it Chandra} and {\it XMM-Newton} much more detailed studies of
ionized absorbers have become possible.  X-ray spectra obtained
with the Reflection Grating Spectrograph (den Herder et al 2001) on
board {\it XMM-Newton} have shown that the absorbing gas has often
various components with different outflowing velocities and clearly
distinct ionisation states (see Sako et al 2001 for IRAS 13349+2438
and Blustin et al 2002 for NGC 3783).  One of the most remarkable
features discovered in these spectra is the presence of Unresolved
Transition Arrays (UTAs) of Fe M lines, which could mimic the shape of
absorption edges when observed at the lower spectral resolution
typical of CCDs. 

Detailed studies of particular AGN which exhibit X-ray ionized
absorption features also demonstrate that there is frequently a narrow-line
absorber related to the AGN (i.e. an {\it associated} absorber) in the
ultraviolet band, normally showing up as a CIV $\lambda\lambda 1548,1550$
absorption feature (Mathur 1994, Mathur et al 1994, Mathur, Elvis \&
Wilkes 1995).  The same photoionized outflowing gas can provide the UV
absorption lines and the ionized absorption features seen in X-ray
spectra.  More recently, Crenshaw et al (1999) found in
a systematic study of UV absorption lines towards nearby radio-quiet
AGN, that {\it all} objects showing UV associated absorption systems
have a corresponding warm absorber in the X-rays.

Variability has also been found in some of these warm absorbers.  Fabian
et al (1994) reported different absorption optical depths in the
best-studied warm absorber towards MCG-6-30-15. Later, Otani et al
(1996) showed, using a 4-day long ASCA observation of this particular
Seyfert 1, that the OVIII K edge (at a rest frame energy of 0.871 keV)
varied on scales of $\sim 10\, {\rm ks}$, while the OVII K-edge (at a
rest frame energy of 0.739 keV) remained constant.  Otani et al (1996)
interpreted this in terms of two different ionized absorbers, a high
ionisation (OVIII) absorber located within the BLR and a lower
ionisation absorber (OVII) probably associated to the Narrow Line
Region (NLR).

H1419+480 was discovered in the Modulation Collimator - Large Area Sky
Survey (MC-LASS) conducted with the HEAO-1 observatory (Wood et al
1984).  The inferred 2-10 keV flux was $\sim 2\times 10^{-11}\, {\rm
erg}\, {\rm cm}^{-2}\, {\rm s}^{-1}$, assuming a power law spectrum
with a canonical $\Gamma=1.7$ (Ceballos \& Barcons 1996). Although the
modulation collimator helped to pin down the position of the X-ray
source, the flux could be severely affected by the contribution of
unresolved sources within the collimator field of view.  The $ROSAT$
All-Sky Survey (RASS) also detected this source (see Schwope et al
2000), with a 0.5-2 keV flux (corrected for Galactic absorption) of
$\sim 7 \times 10^{-12}\, \uflux$ and an additional PSPC pointed
observation conducted in 1992 found a flux of $\sim 3\times 10^{-12}\,
\uflux$. In both cases the PSPC hardness ratio was $\sim 0$ which
excluded a significant photoelectric absorption by cold gas.

Remillard et al (1993) identified  H1419+480 with a broad-line
type 1 AGN at $z=0.072$ ($RA=14^h21^m29.60^s$, $DEC=+47^{\circ}47'27''$).
Appendix A presents our own optical spectroscopic observations of
H1419+480, which confirm that this source is a broad-line AGN with
$z=0.07229$, as derived from the [OIII] emission doublet.

Ceballos \& Barcons (1996) analyzed a sample of sources detected both
in the MC-LASS sample and by $ROSAT$ (which included H1419+480). The
large decrement of the flux from hard to soft X-rays is suggestive of
heavy photoelectric absorption, which is at odds with the value
reported for the PSPC hardness ratio.  This was interpreted by Ceballos
\& Barcons (1996) as evidence for complex absorption, that could be
modelled either in terms of an ionized absorber or in terms of a
partial covering cold absorber.  

In this paper we present {\it XMM-Newton} EPIC observations of
H1419+480. We find that the source has changed the soft X-ray flux
again and that its 2-10 keV flux is almost a factor 3 lower than the
HEAO-1 one.  A warm absorber is unambiguously seen in the {\it
XMM-Newton} data, through mostly OVII absorption, but we argue that
this warm absorber was weaker or absent in the previous pointed
$ROSAT$ observation.  An associated CIV absorption line is seen in an
{\it International Ultraviolet Explorer (IUE)} spectrum of this source
taken more than 20 years ago, but it is not seen in a similar
observation taken one year later.  We find that both X-ray and UV
absorption can be explained in terms of the same amount of gas but
varying ionising conditions.

\section{X-ray properties}

\subsection{{\it XMM-Newton} observations}

H1419+480 was observed by {\it XMM-Newton} (Jansen et al 2001) for 27 ks on
the 27th of May of 2002, during revolution 451 (obsid=0094740201), as
part of the Guaranteed Time programme of the Survey Science Centre. In
this paper we analyze only the data obtained by the EPIC MOS (Turner
et al 2001) and EPIC pn (Str\"uder et al 2001) cameras.  The source is
clearly detected by the RGS spectrographs, but
it is far too faint to deliver any scientifically interesting results.
The OM (Mason et al 2001) was used to obtain images in the filters U,
B and UVW1.

All 3 EPIC cameras had the 'Thin 1' filter on, and they were operated
in partial window mode (MOS1 and MOS2) and small window mode
(pn). Exposure times (excluding overheads) were 13 ks for the MOS
cameras and 20 ks for the pn camera. However, most of the exposure
time was lost due to high background flaring.   After cleaning out
these intervals, good-time intervals of $\sim 8$ ks were left for the
MOS1 and MOS2 detectors, and 3 ks were left for the pn
detector. Fortunately H1419+480 is bright enough (5 cts/s in the EPIC
pn camera) to provide a large enough number of counts for spectral analysis.

Event files for the 3 EPIC detectors were taken from the distributed
pipeline products, which were obtained by processing the observation
data file with SAS v5.3.3.  We filtered out high-background intervals,
keeping only the most reliable single and double events and, in the
case of EPIC pn, by removing those with low spectral quality.
Calibration matrices were generated by using SAS v5.4.1, which we
found to solve some problems at energies 0.5-1 keV with respect to the
SAS v5.3.3 calibration.

\begin{figure}
\includegraphics[height=8cm,angle=270]{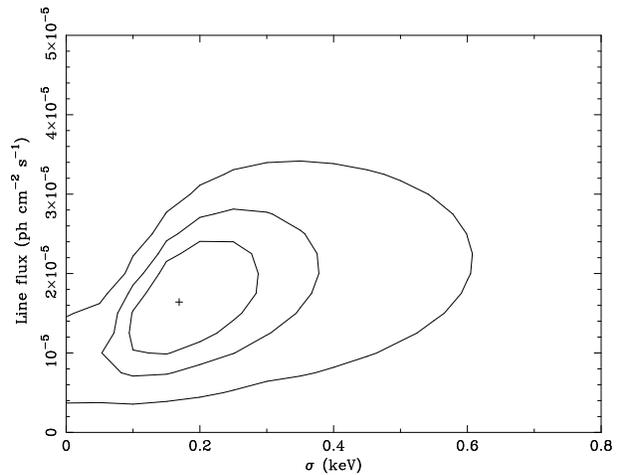}
\caption{Confidence contours (1,2 and 3 sigma) for the flux versus
  dispersion of the Fe line}
\label{contfeline}
\end{figure}

X-ray spectra for H1419+480 and background, were extracted from the 3
EPIC cameras.  The background spectrum was extracted from regions in
the same chip as the source but not illuminated by H1419+480. This
was also done in the MOS data, as opposed to source-free regions in
the outer chips, to prevent vignetting affecting the background
subtraction. The source plus background pn, MOS1 and MOS2 spectra were
binned individually into bins containing at least 20 counts each. To
fit spectral models to these data, {\tt xspec} version 11.2.0 was used
(Arnaud 1996). Bins outside the 0.2-12 keV bandpass were ignored. At
the time of performing this analysis, a calibration problem (probably
produced for an incorrect redistribution matrix) of the MOS
detectors below 0.5 has been reported. In this paper we have taken
the conservative approach of ignoring all MOS data below 0.5 keV.

\subsection{The 2-12 keV X-ray spectrum}

We first fit the 2-12 keV spectrum to determine the underlying
continuum.  A single power law fit gave $\chi^2=305.96$ for 311
degrees of freedom (two parameters fitted), but with obvious residuals
around 6 keV.  Adding a redshifted gaussian emission line improved the
fit substantially to $\chi^2=290.17$ for 308 degrees of freedom.  The
F-test assigned a significance to the detection of this line of
99.90\%. Parameters resulting from the fit are listed in
table~\ref{Xraypars}, while Fig.~\ref{contfeline} shows the confidence
contours in the line flux versus line dispersion parameter space.  The
line equivalent width derived from these parameters is
$240^{+125}_{-110}\, {\rm eV}$. All parameters are entirely consistent
with those from other Seyfert 1 galaxies (Nandra \& Pounds 1994).  The
EPIC pn spectrum shows a hint of a residual towards low energies,
reminiscent of relativistic effects in a disk line, but since the MOS1
and MOS2 spectra do not show such resudials we do not explore this
point any further.  This emphasizes again the importance of
calibration in the detection of weak features (such as relativistic
line profiles) in moderate signal-to-noise data. In what follows the
energy of the Fe line and its width have been fixed to their best-fit
values (see Tab.~\ref{Xraypars}) in order to prevent runaway solutions
with unphyiscal parameters (e.g., extremely broad lines).

\begin{table}
\caption{X--ray spectral parameters of H1419+480 for the 2-12 keV
  fit. All errors are 90\% confidence for 1 interesting parameter.}
\label{Xraypars}
\begin{center}
\begin{tabular}{l c c}
Parameter & Value  & Units\\
\hline
{\tt zgauss} & ($z=0.07229$)\\
$E_{line}$ & $6.52^{+0.12}_{-0.11}$ & keV \\
$\sigma_{line}$ & $0.17^{+0.14}_{-0.06}$ & keV\\
$F_{line}$ & $(1.7^{+0.8}_{-0.8})\times 10^{-5}$ & ${\rm ph}\, {\rm
  cm}^{-2}\, {\rm s}^{-1}$\\
\hline
{\tt zopwerlaw}& ($z=0.07229$)\\
$\Gamma$ & $1.84^{+0.02}_{-0.01}$ & \\
$A_{\Gamma}$ & $(2.53\pm 0.03)\times 10^{-3}$ & ${\rm ph}\, {\rm
  cm}^{-2}\, {\rm s}^{-1}\, {\rm keV}^{-1}$\\
\hline
\end{tabular}
\end{center}
\end{table}

\subsection{The 0.2-2 keV X-ray spectrum}

\begin{figure}
\includegraphics[height=8cm,angle=270]{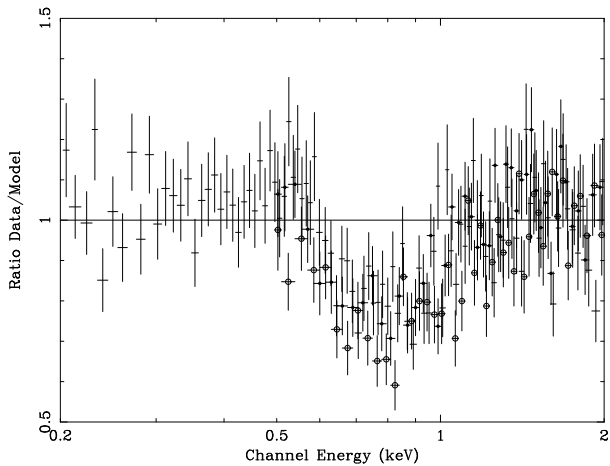}
\caption{{\it XMM-Newton} EPIC spectral ratio of data to a power law
  plus galactic absorption model, where the 0.5-1 keV region has been
  excluded from the fit. Crosses are from EPIC pn,
  filled circles from EPIC MOS1 and hollow circles from EPIC MOS2. The
  data points have been binned into 100 count bins for illustration
  purposes only.}
\label{ratio}
\end{figure}

We then extend the fit to 0.2 keV, including photoelectric absorption
from the Galaxy ($N_H=1.65\times 10^{20}\, {\rm cm}^{-2}$).  The
fit is not good, with $\chi^2=1140.45$ for 795 degrees of freedom.
There are obvious negative residuals as shown in fig.~\ref{ratio},
where the 0.5-1 keV band has been excluded from the fitting region. 
We have explored a number of model fits in trying to understand this
feature.  

\subsubsection{Absorption plus soft excess}

We first try to fit the spectrum with an absorbed power law plus a
soft excess in trying to mimic the negative feature seen in the
residuals.  The soft excess is modelled as a steep power law (a black
body model did not converge to a better value of the $\chi^2$).  The
best fit yields a minimal reduced $\chi^2_{\nu}=936.03/792$,
corresponding to a $\Gamma_{\rm soft}=2.40 \pm 0.09$ and to an
intrinsic absorption (at the redshift of the Seyfert 1 $z=0.07229$) of
$N_H=(5.7_{-1.0}^{+1.2})\times 10^{21}\ {\rm cm}^{-2}$.  Although the
improvement in the $\chi^2$ is very significant, strongly correlated
negative residuals in the 0.5-1 keV band persist.  One of the models
suggested by Ceballos \& Barcons (1996) to explain the unusual X-ray
colours of this source on the basis of a partial covering cold
absorber, yields an even worst $\chi^2=983.78$.

\subsubsection{Single ion absorption}

The absorption trough seen in the residuals is clearly reminiscent of
an absorption edge or a UTA.  The ``equivalent width'' of the
absorption feature seen in the data is $\sim 0.7$~\AA, which is twice
larger than the values typically found in the Fe M UTAs. Besides,
the absorption feature in H1419+480 spreads over $\sim 5$~\AA, while
the Fe M UTAs spread over $1.5$~\AA\ in the RGS.  The spectral
resolution of EPIC at the energies of interest is $0.6-1$~\AA, and
therefore would be unable to spread a UTA up to 5\AA. A further
argument against the bulk of this feature being due to a Fe M UTA
comes from the fit to a multi-ion photoinised absorber (see subsection
2.3.3) where all Fe M lines are included and yet the absorber's
parameters are very similar to those fitted by assigning the feature
to (mostly) the OVII absorption edge.  Therefore we conclude that,
although UTAs can contribute to the detected absorption feature, they
cannot account for all of it. We therefore start by attempting
to interpret the negative spectral feature in terms of an absorption
edge.

Indeed, adding a single absorption redshifted edge to the data results
in a minimal $\chi^2_{\nu}=854.11/793$ (i.e., substantially better
than the soft excess model). The edge energy is
$E_{edge}=715_{-19}^{+10}\, {\rm eV}$ with a maximum depth of
$\tau_{edge}=0.58_{-0.05}^{+0.07}$. If interpreted as an intrinsic
OVII absorption K-edge, at a rest frame energy of 739 eV, the absorber
should be infalling towards H1419+480 at a velocity
$v=9100_{-3800}^{+7200}\, {\rm km}\, {\rm s}^{-1}$.  This velocity
shift is significantly larger than the velocity outflow found in the
best studied cases with the X-ray gratings.

\begin{figure}
\includegraphics[height=8cm,angle=270]{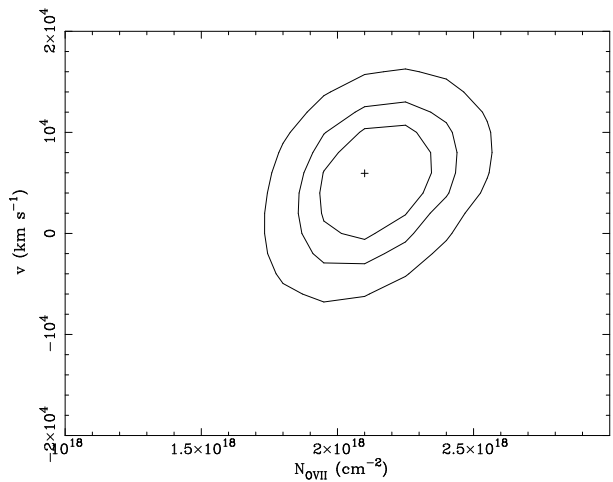}
\caption{Best fit and confidence contours (1,2, and 3 sigma) for the
  parameters in the single ion absorption model {\tt siabs} to the
  absorption feature.}
\label{cont_siabs}
\end{figure}

One fact that might contribute, at least partly, to this shift is the
presence of resonance absorption lines from the same ion, which are
unresolved by EPIC, but still contributing to the shape of the
absorption feature.  This fact has been noted, among others, by Lee et
al (2001) in the analysis of the {\it Chandra} data of the Seyfert 1
galaxy MCG-6-30-15. We therefore prefer to model the single ion
absorption by the {\tt siabs} model released in the PHOTOION package
(Kinkhabwala et al 2003), under the assumption that the feature
corresponds to OVII. The fit results in a $\chi^2_{\nu}=851.41/793$
corresponding to a column density $N_{OVII}=(2.1\pm 0.3)\times
10^{18}\, {\rm cm}^{-2}$ and an infall velocity of
$v=6000_{-7350}^{+5400}$ towards the source.
Fig.~\ref{cont_siabs} shows the confidence contours for these two
parameters. We note that the fit is even marginally better than the
one provided by the K-edge solely and that the velocity shift is now
much more moderate (and consistent with zero), as in better-studied
warm absorbers.

\subsubsection{Multiple-ion absorption}

\begin{figure}
\includegraphics[height=8cm,angle=270]{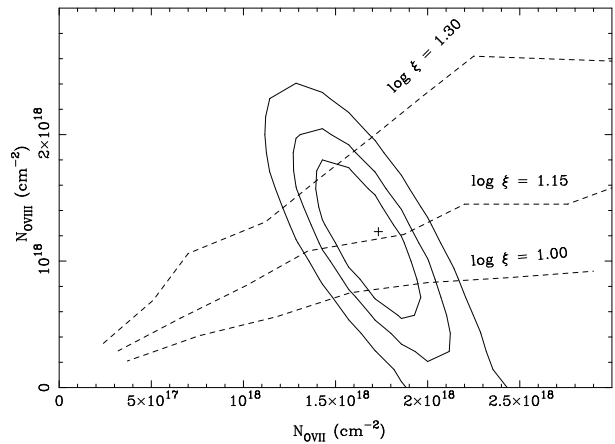}
\caption{Best fit and confidence contours (1,2, and 3 sigma) for the
  OVII and OVIII column densities by fitting 2 single ion absorption
  with infall velocity fixed at $6000\, {\rm km}\, {\rm s}^{-1}$.
  The dashed lines show the predictions for a thermal ($T=3\times
  10^5\, {\rm K}$) photoionised gas and representative values of the
  ionisation parameter $\xi$, for a range of H column densities. }
\label{cont_2_siabs}
\end{figure}

\begin{figure}
\includegraphics[height=8cm,angle=270]{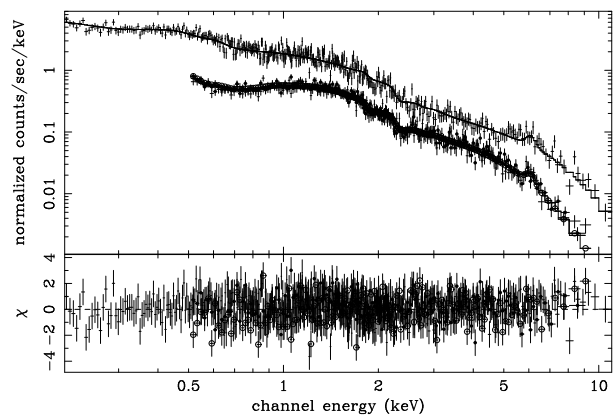}
\caption{{\it XMM-Newton} EPIC spectrum and residuals with respect to the
  best-fit model consisting in a power law plus Fe line and 
  absorption for both OVII and OVIII. Crosses are from EPIC pn,
  filled circles from EPIC MOS1 and hollow circles from EPIC MOS2.}
\label{Xrayspec}
\end{figure}

Although the single ion absorption fit does not leave obvious
residuals in the X-ray data, we have attempted to add an OVIII
absorber with the same recession  velocity than the OVII absorber
fixed at the best fit value of $6000\, {\rm km}\, {\rm s}^{-1}$.  The
$\chi^2_{\nu}$ further decreases to $846.68/792$ by just adding one parameter
(the OVIII column density). The F-test statistic yields a significance
of only 96.4\% for this new component.  Fig~\ref{cont_2_siabs} shows
the confidence contours in the OVII and OVIII column density parameter
space.   Fig.~\ref{Xrayspec} shows the EPIC spectrum together with the 2 single
ion absorption spectral fit, along with the featureless residuals.

\begin{figure}
\includegraphics[height=8cm,angle=270]{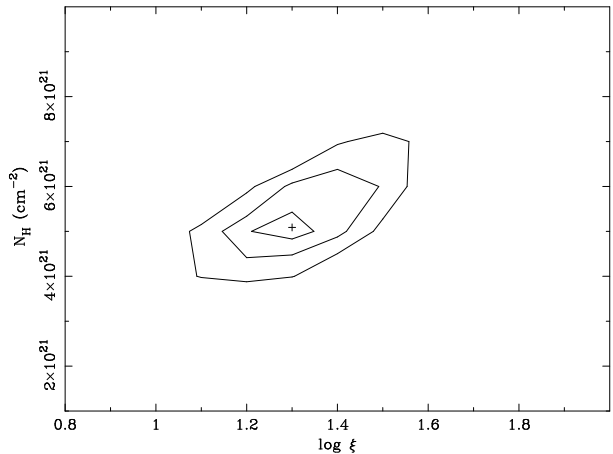}
\caption{Best fit and confidence contours (1,2, and 3 sigma) for the
  total column density and ionisation parameter $\xi$ (in units of
  ${\rm erg}\, {\rm cm}\, {\rm s}^{-1}$) after fitting the X-ray
  spectrum to a photoionised gas with solar abundances (model {\tt
  xiabs})} 
\label{cont_xiabs}
\end{figure}

We have also explored the possibility of describing the absorption
feature in terms of a solar abundance photoinized gas, by using the
model {\tt xiabs} from the PHOTOION package (Kinkhabwala et al 2003).
The best fit, by fixing the infall velocity at $6000\, {\rm km}\, {\rm
  s}^{-1}$, is reached for $\log\xi=1.30\pm 0.1$ and a total H column
density of $N_H\sim (5.1\pm 0.7)\times 10^{21}\, {\rm cm}^{-2}$ with a
$\chi^2_{\nu}=871.32/793$. Contour plots are shown in
Fig.~\ref{cont_xiabs}. However the fit is not as good as that for the
OVII and OVIII absorption only, presumably due to unmatched features
from other elements.

The {\tt xiabs} model used does not include thermal (collisional)
ionisation in the absorbing gas.  We illustrate the effects of
collisional ionisation by computing the values of $N_{OVII}$ and
$N_{OVIII}$ for a photoionised gas at a temperature $T=3\times 10^5\,
{\rm K}$ (typical of warm absorber models) and assuming a gas density
of $10^{10}\, {\rm cm}^{-3}$ (e.g., as in the Mathur et al 1994
models), although this last parameter is only marginally
relevant as the gas is optically thin. We have used XSTAR (version
2.1) along with solar abundances and the results are shown in
Fig.~\ref{cont_2_siabs} for a few values of the ionisation parameter.
In this case, some of the ionisation takes place by collisions and
therefore the required ionisation parameter is marginally smaller
($\log\xi\sim 1.15$).  The required value for the column density is
also around $N_H\sim 5\times 10^{21}\, {\rm cm}^{-2}$. With this we
conclude that both the ionisation parameter and the amount of gas
along the line of sight are fairly well established.

\subsection{Variability}

Although the {\it XMM-Newton} exposure is relatively short, we have
also explored whether the source varied within the observation.
Counts have been accumulated in the 0.5-2 keV and 2-12 keV band. The
reason for this resides in the fact that the 0.5-2 keV band is
dominated mostly by the absorbed component and the 2-12 keV by the
underlying continuum (plus Fe line). 

\begin{figure}
\includegraphics[height=8cm,angle=270]{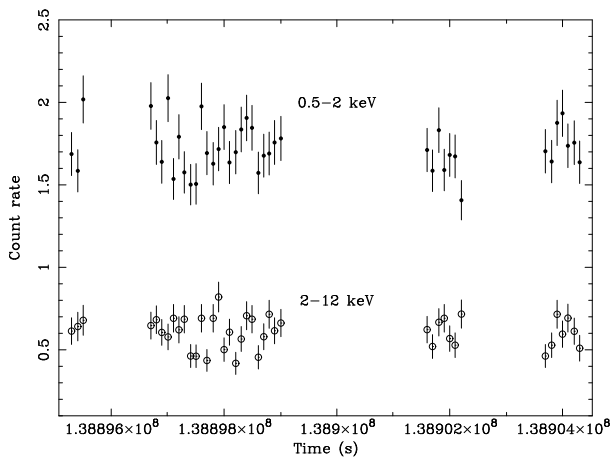}
\caption{EPIC-pn light curves of H1419+480 in the 0.5-2 keV and 2-12
keV bands. The gaps are due to high-background flaring intervals.}
\label{lc}
\end{figure}

Counts have been grouped into 100 s bins, which produced an EPIC pn count rate
of 1.7 ct/s in the 0.5-2 keV band and 0.6 ct/s in the 2-12 keV
band. Although it is a minor correction, the lightcurve of the
background has been scaled and subtracted from the source's
lightcurve. Fig.~\ref{lc} shows the resulting lightcurves of
H1419+480. In order to look for the signficance of any variability, we
fit a constant value to every lightcurve and compute the minimum
$\chi^2$.  For the 0.5-2 keV band, $\chi^2=48.58$ for 41 points, which
yields a probability of the source being variable of only 83\%.  In
the 2-12 keV band $\chi^2=60.9$, with a probability of the source
being variable of 98\%.  We have further checked whether any marginal
variations are different in both bands, by fitting a constant to the
difference between the 2-12 keV and 0.5-2 keV count rates.  The
$\chi^2$ fit to a constant yields $\chi^2=32.0$ for 40 points, which
is entirely consistent with no X-ray colour variations.  We therefore
conclude that there is no significant variability of the source within
the {\it XMM-Newton} observation and that in any case variations
preserve the X-ray colours of the source. Therefore there are no
constraints on the physical models for this source derived from
variability.

\subsection{Comparison to previous X-ray observations}

Using the best-fit model with two single ion absorption (OVII and
OVIII), we compute the flux of H1419+480, corrected for Galactic
absorption, to be $4.5\times 10^{-12}\, {\rm erg}\, {\rm cm}^{-2}\,
{\rm s}^{-1}$ in the 0.5-2 keV band and $7.3\times 10^{-12}\, {\rm
erg}\, {\rm cm}^{-2}\, {\rm s}^{-1}$ in the 2-10 keV band.  Adopting
currently fashionable cosmological parameters ($H_0=70\, {\rm km}\,
{\rm s}^{-1}\, {\rm Mpc}^{-1}$, $\Omega_m=0.3$ and
$\Omega_{\Lambda}=0.7$), the luminosity of H1419+480 is $5.7\times
10^{43}\, {\rm erg}\, {\rm s}^{-1}$ in the 0.5-2 keV band and
$9.2\times 10^{43}\, {\rm erg}\, {\rm s}^{-1}$ in the 2-10 keV band.

The {\it XMM-Newton} 2-10 keV flux is almost a factor of 3 smaller than the
HEAO-1 flux, a fact that can be at least in part explained by source
confusion in the HEAO-1 collimators (see Barcons et al 2003 for a
discussion on a similar situation for another source).

There are two $ROSAT$ observations of H1419+480 recorded from this
source: one from the $RASS$ (reported in Schwope et al 2000) and a
second one performed on January 1992 (observation sequence ROR700038),
used in Barcons \& Ceballos (1996).  Table~\ref{ROSAT} lists some data
of these observations, along with expectations from the {\it XMM-Newton}
observation.  We have folded our best-fit model to the {\it XMM-Newton} data
through the $ROSAT$ PSPC-B response to compute the expected spectral
shape in the $ROSAT$ observations and to convert count rates to
fluxes.

\begin{table*}
\caption{Details of $ROSAT$ observations of H1419+480, along with
  expected values of the PSPC Hardness Ratio and Softness ratios
  derived from the best-fit to the {\it XMM-Newton} data with and without
  the OVII+OVIII absorption.}
\label{ROSAT}
\begin{center}
\begin{tabular}{l l l l l}
Parameter  & $ROSAT$ & $ROSAT$ & {\it XMM-Newton} & {\it XMM-Newton}\\
           & (RASS)& (ROR700038) & (with absorption) & (without absorption)\\
\hline
Exposure time (s) & 375 & 1266 & & \\
Flux$^a$ & $7\times 10^{-12}$ & $3\times 10^{-12}$ & $4.5\times
10^{-12}$ & $4.5\times 10^{-12}$\\
$HR^b$ & $+0.0\pm 0.1$ & - & +0.0 & +0.1\\
$SR_1^c$ & - & $+1.6\pm 0.2$ & +2.4 & +1.7\\
$SR_2^d$ & - & $+0.7\pm 0.1$ & +0.6 & +0.7\\
\hline
\end{tabular}
\end{center}
\hbox{$^a$ Flux in the 0.5-2 keV band in units of ${\rm erg}\, {\rm
  cm}^{-2}\, {\rm s}^{-1}$, corrected for Galactic absorption
using the best fit model from table \ref{Xraypars}}
\hbox{$^b$ PSPC Hardness Ratio $HR=(H-S)/(H+S)$ where $S$ and $H$ are the
  counts detected in PSPC channels 11-39 and 50-200 respectively.}
\hbox{$^c$ PSPC softness ratio $SR_1=S_1/H_1$, where $S_1$ and $H_1$ are the
  counts detected in PSPC channels 11-39 and 40-85 respectively.} 
\hbox{$^d$ PSPC softness ratio $SR_2=H_1/H_2$, where $H_2$ are the
  counts detected in PSPC channels 86-200.} 
\end{table*}

The first obvious conclusion is that the 0.5-2 keV flux has changed
significantly between the 3 different observations.  The expected
spectral shape in the $ROSAT$ observations appears similar (but not
identical) to the measured one. The PSPC Hardness and Softness Ratios
(see caption of table~\ref{ROSAT} for definitions) expected in the
$ROSAT$ data from our best-fit model to the {\it XMM-Newton} spectrum
have been computed with and without the OVII and OVIII absorption
components.  In particular the value measured for $SR_1$ in the 1992
observation appears consistent with the absence (or weakening) of the
absorption features, but largely inconsistent with the presence of the
absorption features.  We conclude that the absorption features were
much weaker or absent during the $ROSAT$ observation of H1419+480 in
1992.

\section{An associated CIV absorber}

The compelling evidence that X-ray warm absorbers share a common
origin with associated narrow-line absorbers in the UV (Crenshaw et al
1999) urged us to search in available archives for UV spectroscopic
observations of H1419+480.  Unfortunately the {\it Hubble Space
Telescope (HST)} has not observed this object with any of the UV
spectrographs, but the {\it IUE} did
observe it several times. Table \ref{IUE} lists the observations
conducted with the SWP short wavelength camera (operated at low
dispersion $\sim 6$\AA), which encompass both
the Ly$\alpha$ $\lambda 1216$ and the CIV $\lambda 1549$ emission lines.
We analyzed the INES ({\it IUE} Newly Extracted Spectra) data from both of
these observations.

\begin{figure}
\includegraphics[height=8cm,angle=270]{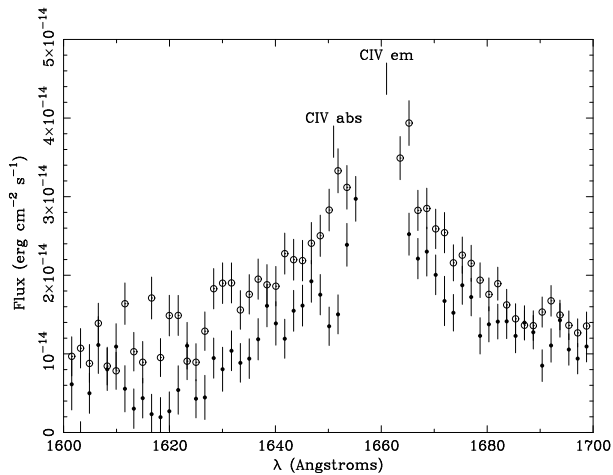}
\caption{{\it IUE}/SWP spectra of H1419+480 around the CIV emission line.
  Only datapoints which have not been flagged by any reason are
  plotted. Filled (hollow) circles are from the SWP17265 (SWP18951)
  observations respectively.}
\label{IUEspec}
\end{figure}

Fig.~\ref{IUEspec} shows the spectra obtained by {\it IUE} during both
observations, around the CIV emission line. Only channels which have
not been flagged for any reason are plotted. The overall flux shift
between both spectra could be either true variability or a calibration
problem. The most obvious discrepancy between both spectra occurs at
around $\lambda\sim 1651$\AA, where the earlier observation (SWP17265)
shows a dip that could be an absorption line.

To further investigate this, and keeping in mind that detecting
absorption lines on top of emission lines is a difficult task, we have
taken the 8 channels ranging from 1643 to 1656 \AA\ in the SWP17265
spectrum. A linear function was fitted to that range, resulting in a
$\chi^2=17.65$ for 6 degrees of freedom.  Multiplying by a  gaussian
absorption line and freezing its width to the resolution of the
spectrograph (i.e., searching for an unresolved line) the $\chi^2$
improved to $\chi^2=0.99$ for 4 degrees of freedom.  The F-test
significance of that feature is 99.7\% ($\sim 3\sigma$). The
central wavelength is $\lambda_{abs}=1651.1\pm 0.5$\AA, corresponding
to an outflowing velocity of $\sim 1800\pm 90\, {\rm km}\, {\rm
s}^{-1}$. The equivalent width is also very uncertain, due to the poor
sampling, but formally the best fit is $3^{+0.7}_{-0.9}$\AA. 

\begin{figure}
\includegraphics[width=8cm]{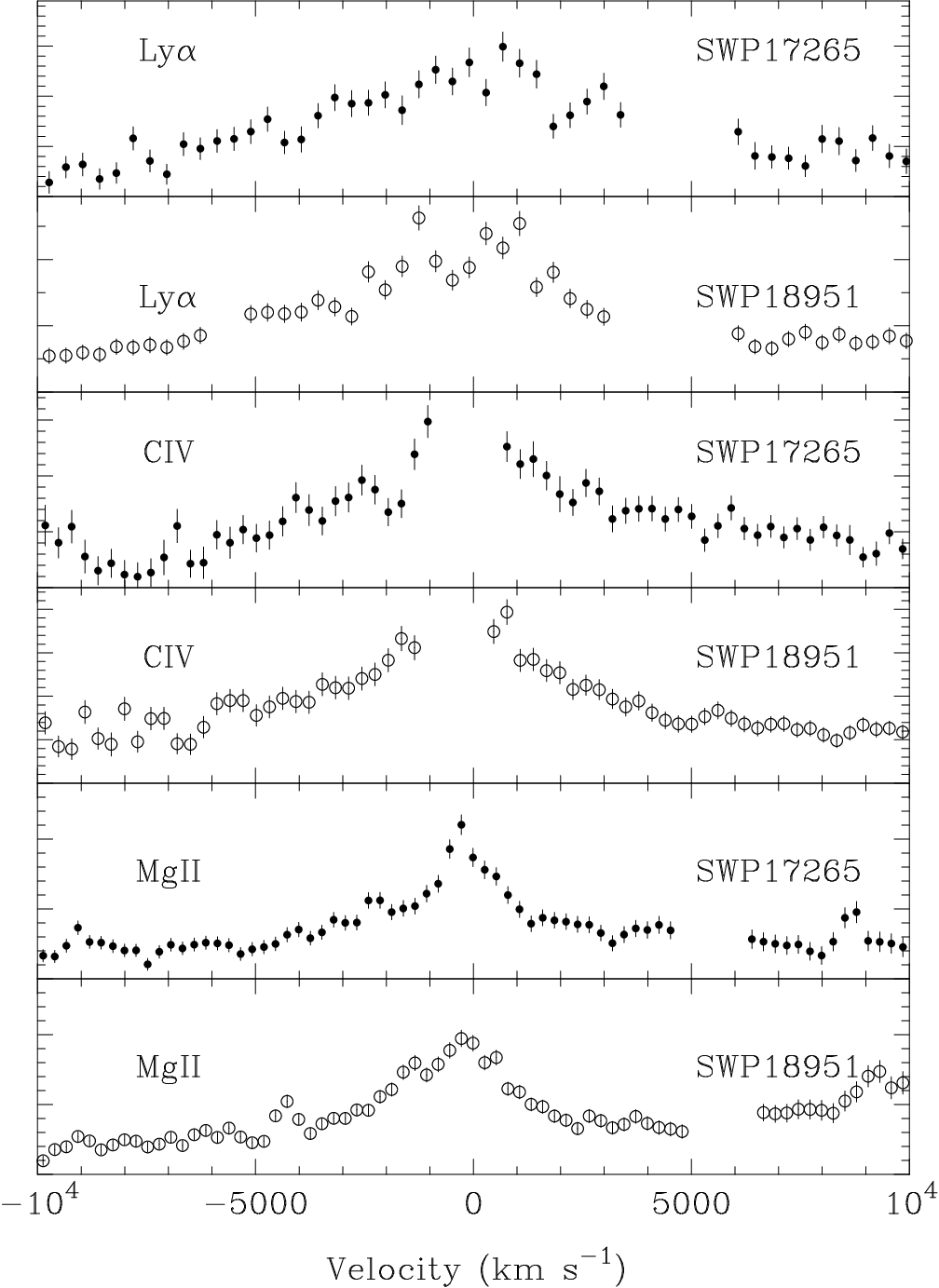}
\caption{{\it IUE}/SWP velocity spectra of H1419+480 around the Ly$\alpha$,
  CIV and MgII emission lines.
  Only datapoints which have not been flagged by any reason are
  plotted. Filled (hollow) circles are from the SWP17265 (SWP18951)
  observations respectively.}
\label{UVlines}
\end{figure}

We have also explored the wings of the Ly$\alpha$ $\lambda 1216$ and
MgII $\lambda 2800$ (see fig.~\ref{UVlines}), but no other obvious
absorption lines are evident. We conservatively estimate that a
3$\sigma$ upper limit for any Ly$\alpha$ absorption line present in
this spectrum is about $3$\AA.

\begin{table}
\caption{Details of {\it IUE}/SWP observations of H1419+480.}
\label{IUE}
\begin{center}
\begin{tabular}{l l c}
Image identification & Date  & Exposure (s)\\
\hline
SWP17265 & 1982-06-19 & 8400\\
SWP18951 & 1983-01-05 & 9000\\
\hline
\end{tabular}
\end{center}
\end{table}

\begin{figure}
\includegraphics[height=8cm,angle=270]{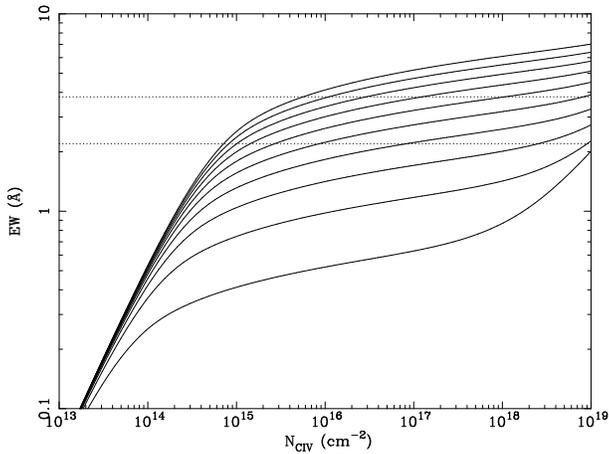}
\caption{Curve of growth for the CIV $\lambda\lambda$1548,1550 doublet
(continuous lines). Velocity dispersion parameters $b=\sqrt{2}\sigma$
($\sigma$ is the velocity dispersion of the gas) range from 20 to 200
${\rm km}\, {\rm s}^{-1}$ in steps of 20 ${\rm km}\, {\rm s}^{-1}$
from bottom to top. The dotted lines show the measured equivalent
width formal 90\% confidence interval.}
\label{cog_civ}
\end{figure}

At the 6\AA\ resolution of this setup, the CIV doublet
($\lambda\lambda 1548, 1550$) is unresolved and, obviously, we cannot
measure directly the absorbing column density. Fig~\ref{cog_civ} shows the
curve of growth (i.e., equivalent width versus column density) for the
whole CIV doublet and a range of velocity dispersion parameters. As
usual a Voigt profile has been assumed for each line in the doublet,
with velocity width parameter $b=\sqrt{2}\sigma$ ($\sigma$ is the
velocity dispersion of the gas, assumed Maxwellian), damping constant
$2.64\times 10^8\, {\rm s}^{-1}$ and a {\it total} oscillator strength
of 0.28 (which is the sum of the oscillator strengths for the two
lines).  The approximation introduced by Whiting (1968) to the Voigt
profile (accurate to better than 5\%) has been used.

\begin{figure}
\includegraphics[height=8cm,angle=270]{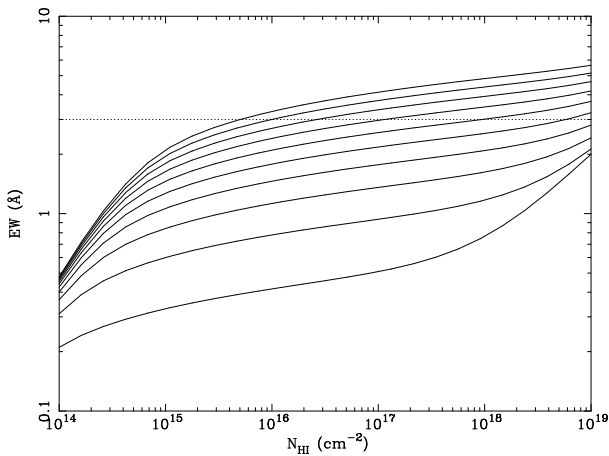}
\caption{Curve of growth for the HI Ly$\alpha$ $\lambda$1216
transition (continuous lines). Velocity dispersion parameters as in
Fig.~\ref{cog_civ}. The dotted line shows the adopted 3$\sigma$
upper limit for the detection of this line in the {\it IUE} spectra.}
\label{cog_lya}
\end{figure}

The amount of CIV absorbing gas is very uncertain, as the measured
equivalent width falls in the saturated part of the curve of growth,
but values around $N_{CIV}\sim 10^{15-16}\, {\rm cm}^{-2}$ would be
appropriate for Doppler velocity parameters around $100\, {\rm km}\,
{\rm s}^{-1}$. In Fig.~\ref{cog_lya} we also plot the curve of growth
for HI Ly$\alpha$, using an oscillator strength of 0.416 and a damping
constant of $4.7\times 10^8\, {\rm s}^{-1}$, along with the upper
limit of $3$\AA.  We can see that relatively large HI column densities
(up to $10^{17-18}\, {\rm cm}^{-2}$) could go undetected for similar
values of the Doppler velocity parameter.

\section {Discussion} \label{theory}

Establishing a link between the X-ray and the UV absorbing gas is
difficult in this case, as the observations are not simultaneous.  We
can, however, test whether or not it is plausible that the same or a
similar amount of gas (perhaps in a different ionisation state)
produced both absorption features.

The X-ray absorbing gas has a fairly well determined column of
$N_H\sim 5\times 10^{21}\, {\rm cm}^{-2}$ and it is highly ionized
($\log\xi\sim 1.1-1.3$).  Its velocity with respect to the emission
redshift is only poorly defined by the EPIC data but consistent with
zero.  The absorber was significantly weaker or absent in the 1992
$ROSAT$ observation.

The UV gas detected as a CIV associated absorber in the 1982
{\it IUE} observation of H1419+480,  has a
fairly well determined outflowing velocity of $1800\pm 90\, {\rm km}\,
{\rm s}^{-1}$, but its CIV column density is only poorly constrained
$\sim 10^{15}\, N_{15}\, {\rm cm}^{-2}$ with $N_{15}\sim 1-10$.  No HI
Ly$\alpha$ of MgII absorption are found, although the {\it IUE} spectra at
the Ly$\alpha$ region are noisier than around the CIV and MgII
lines. The total column density implied by the CIV absorption is

\begin{equation}
N_H=2.4\times 10^{18}\, N_{15}\, f_{CIV}^{-1}\, {\rm cm}^{-2}
\end{equation}
where $f_{CIV}$ is the fraction of C which is in $CIV$ ionisation
state. Indeed this is a completely unknown parameter, as we cannot
determine the ionisation parameter of this gas which applies to the
1982 observation from a single UV unresolved absorption feature.  We
must restrict ourselves to check whether a value of $N_H\sim 5\times
10^{21}\, {\rm cm}^{-2}$, as in the X-ray absorber, can be obtained
for reasonable ionisation conditions.

Using XSTAR (version 2.1) we have computed ionization fractions for
CIV ($f_{CIV}$), HI ($f_{HI}$) and MgII ($f_{MgII}$) for a range of
values of $\log\xi\sim 1-1.5$, assuming the gas temperature fixed at
$T=3\times 10^5\, {\rm K}$ and with a spectral shape given by a $\Gamma=1.84$
power law.  As pointed out by, e.g., Mathur et al (1994) this might
result in a very inaccurate shape of the UV ionizing continuum and
instead a more accurate shape should be used.  However, for the
purposes of this illustrative exercise, where the ionising parameter
is essentially unknown, this rough modelling of the ionising spectrum
is good enough.

For this range of parameters, $f_{CIV}$ ($f_{HI}$) decreases from
$7.1\times 10^{-3}$ ($3.4\times 10^{-5}$) for $\log\xi=1.0$ to
$4.4\times 10^{-5}$ ($4.0\times 10^{-6}$) for
$\log\xi=1.5$. $f_{MgII}$ is zero for this range of ionisation. If we
adopt the amount of gas seen in the X-ray absorber ($N_H\sim 5\times
10^{21}\, {\rm cm}^{-2}$), HI Ly$\alpha$ would be difficult to detect
in all the explored range of ionisations, as the HI column density
would be below $10^{17}\, {\rm cm}^{-2}$.  However, $N_{CIV}$ would
range from $1.6\times 10^{16}\, {\rm cm}^{-2}$ for the lower
ionisation states explored down to $\sim 10^{14}\, {\rm cm}^{-2}$ in
the opposite extreme.  In conclusion, there is a range of ionisation
parameters, in fact with values around those measured from the X-ray
absorber ($\log\xi\sim 1.3$), where it is plausible to observe the CIV
absorption line and no HI Ly$\alpha$ or MgII absorptions.  A slightly
higher ionisation brings also the CIV absorption below detection
limits.

\section{Conclusions}

The bright Seyfert 1 galaxy H1419+480 has been shown to be variable in
X-rays by comparing our {\it XMM-Newton} observations with earlier
$ROSAT$ data.  A warm absorber is clearly seen mostly through OVII
absorption, but also with $\sim 2\sigma$ evidence for OVIII
absorption, in the {\it XMM-Newton} data at the emission redshift.
The absorber was weaker or absent in a $ROSAT$ pointed observation
performed in 1992. The absorber contains a gas column of $N_H\sim
5\times 10^{21}\, {\rm cm}^{-2}$, and the ionisation parameter is
around $\log\xi\sim 1.15-1.3$ depending on whether or not ionisation
by collisions is included.

\begin{figure}
\includegraphics[height=8cm,angle=270]{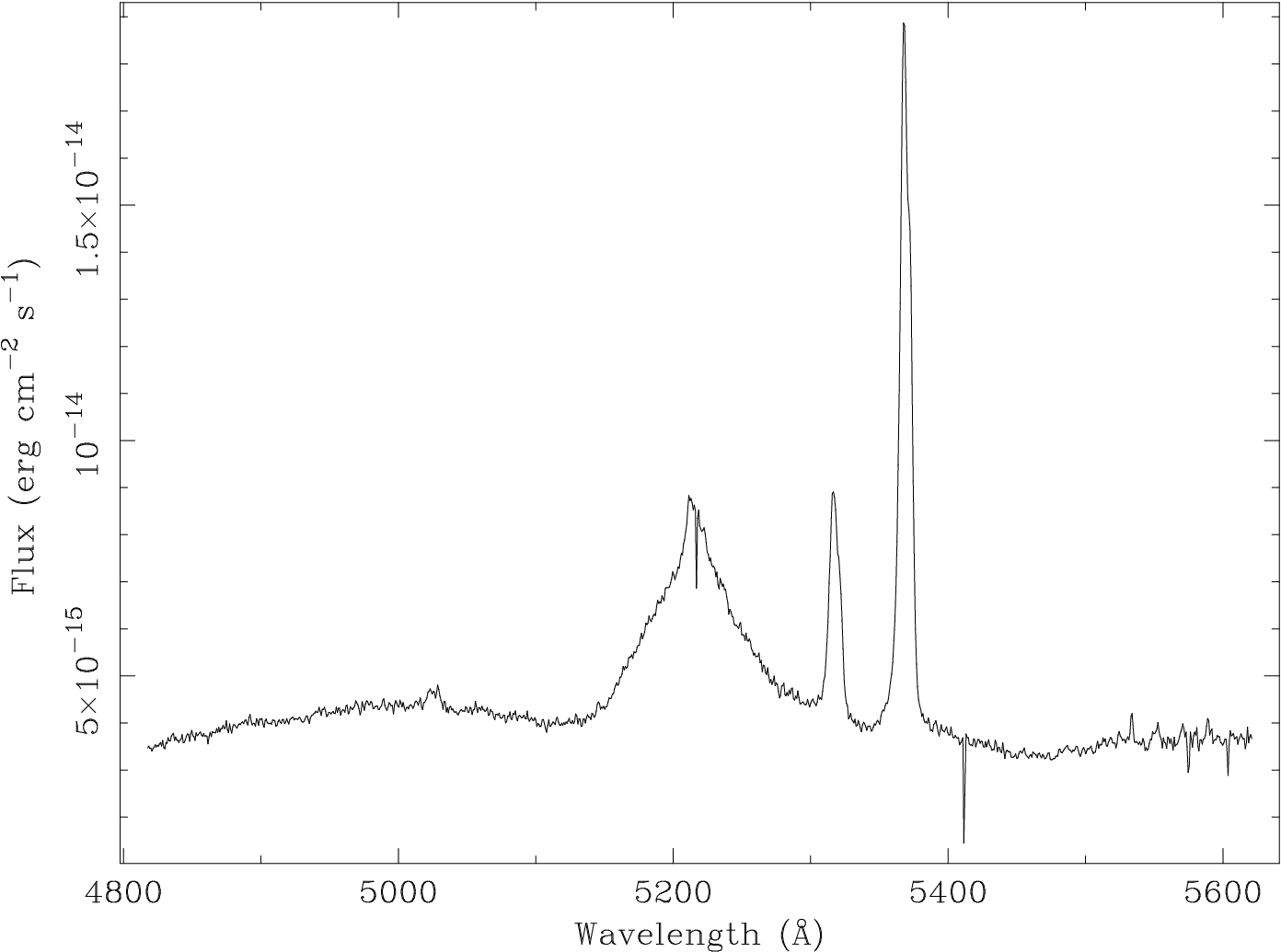}
\includegraphics[height=8cm,angle=270]{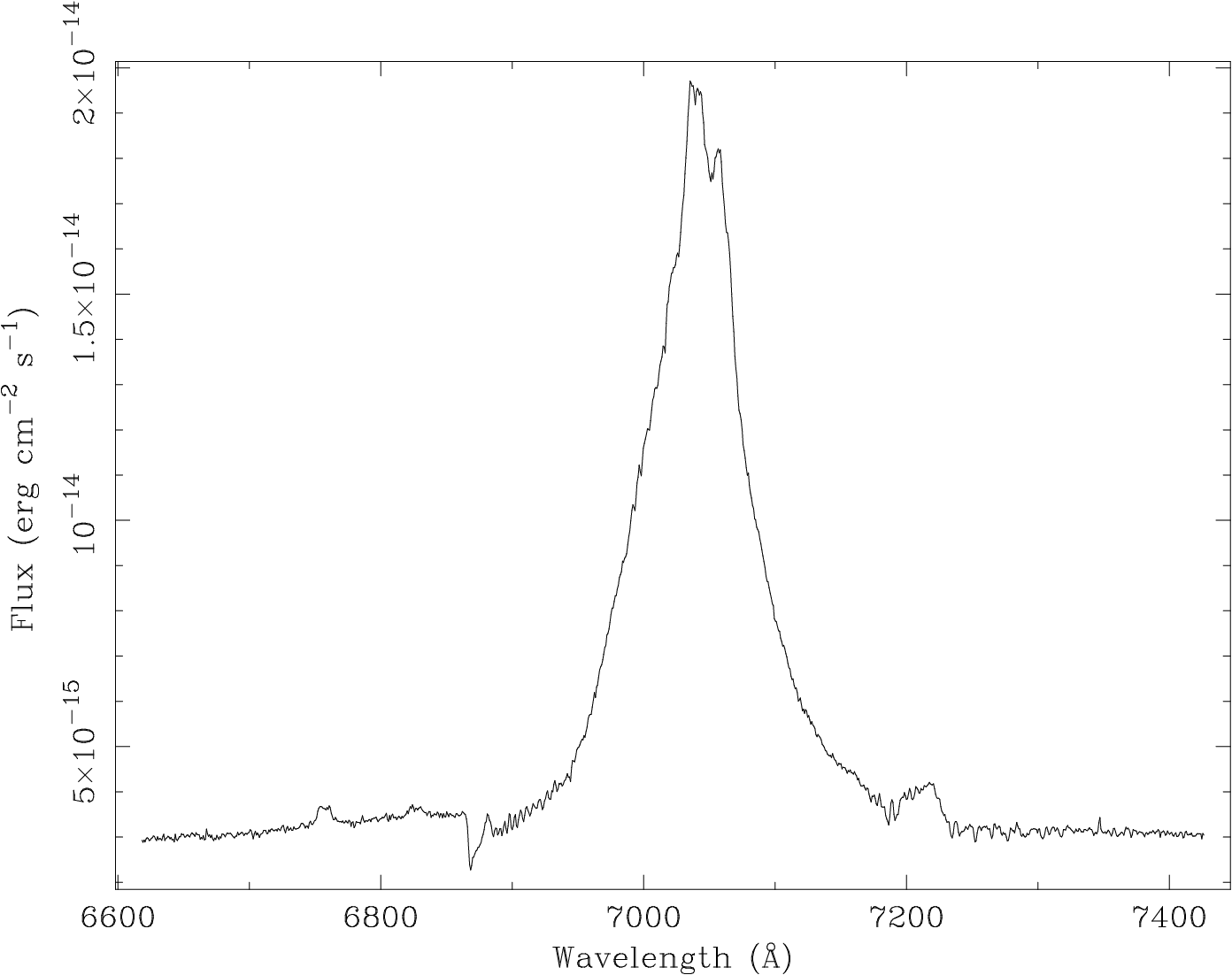}
\caption{Optical spectrum of H1419+480.}
\label{optical}
\end{figure}

An {\it IUE} observation of H1419+480 conducted in 1982 shows $\sim
3\sigma$ detection for a CIV associated absorber, outflowing with a
velocity $\sim 1800\pm 90\, {\rm km}\, {\rm s}^{-1}$.  This confirms
the close link between X-ray ionized absorbers and UV associated
absorption reported, among others, by Crenshaw et al (1999).  This
absorption line was not present in an {\it IUE} observation obtained
roughly 1 year later.

We find that a plausible common origin for all these variable
absorption phenomena can be understood in terms of the same (or
similar) amount of photoionised gas, but with varying ionisation
state. Indeed, for values of the ionisation parameter similar to those
inferred from the X-ray observations, we find that the X-ray absorbing
gas could have produced the CIV absorption feature without any
detectable HI Ly$\alpha$ or MgII lines.  A slightly higher ionisation
would explain the lack of the CIV absorption feature in the {\it IUE}
observation in 1983.

We therefore conclude that H1419+480 has a variable ionised absorber.
Both X-ray and associated UV absorbers are detected at different
epochs in this object, along the lines of the sample studied by
Crenshaw et al. (1999).  Furthermore, we have shown that both
absorbers are plausibly related.


\appendix

\section[]{The optical spectrum of H1419+480}

H1419+480 was observed in the 4.2m William Herschel Telescope at the
Observatorio del Roque de Los Muchachos in the island of La Palma
(Canary Islands, Spain), on February 26, 1998.  We used the ISIS
double spectrograph with 600 line/mm gratings on both the blue and red
arms, with the wavelengths centered at 5200 and 7000 \AA\
respectively, in order to observe the H$\beta$+[OIII] region in the
blue and the H$\alpha$+[NII]+[SII] in the red.  Weather conditions
were good and probably photometric, but the seeing was around 1.5
arcsec. Two 300 sec observations (co-added in the reduction process)
were carried out with the slit aligned to paralactic angle.  For
details on the reduction process, see Barcons et al (2003) as the same
setup and procedure was used.

Fig.~\ref{optical} shows the optical blue and red spectrum of
H1419+480.  The [OIII] lines have some structure, with two
peaks separated $\sim 60\, {\rm km}\, {\rm s}^{-1}$. This is the main
limiting factor in the precission of the redshift, that we measure by
fitting both lines to a common redshift and find $z=0.072296\pm 0.000004$.

\section*{Acknowledgements}
The work reported herein is based partly on observations obtained with
{\it XMM-Newton}, an ESA science mission with instruments and
contributions directly funded by ESA member states and the USA
(NASA). It is also based on INES data from the {\it IUE} observatory. We
thank Enrique Solano for assistance with the INES data and an
anonymous referee for important suggestions. The $WHT$ telescope is
operated by the Isaac Newton Group on the Spanish Observatorio del
Roque de los Muchachos of the Instituto de Astrof\'\i sica de
Canarias.  We acknowledge financial support by the Ministerio de
Ciencia y Tecnolog\'\i a (Spain), under grants AYA2000-1690 and
ESP2001-4537-PE.

\bsp

\label{lastpage}

\end{document}